# Using Annular Josephson Tunnel Junctions to Monitor Causal Horizons

R. Monaco$^a$ and R. J. Rivers$^b$,
a) Istituto di Cibernetica del C.N.R., I-80078, Pozzuoli, Italy
and Unita' INFM-Dipartimento di Fisica,
Universita' di Salerno,
I-84081 Baronissi, Italy
b) Blackett Laboratory, Imperial College London,
London SW7 2AZ, U.K.

(Dated: March 2, 2005)

If systems change as fast as possible as they pass through a phase transition then the initial domain structure is constrained by causality. We shall show how we can trace these causal horizons by measuring the spontaneous production of flux in annular Josephson Tunnel Junctions as a function of the quench time $\tau_Q$ into the superconducting phase. A specific test of our analysis is that the probability $P_1$ to trap a single fluxon at the N-S transition clearly follows an allometric dependence on $\tau_Q$ as $P_1 = a\,\tau_Q^{-\sigma}$, with a scaling exponent $\sigma = 0.25$, in agreement with the data.
PACS Numbers : 11.27.+d, 05.70.Fh, 11.10.Wx, 67.40.Vs

## I. INTRODUCTION

The Kibble-Zurek scenario [1–3] for continuous phase transitions proposes that transitions take effect as fast as possible, i.e. the domain structure initially matches causal horizons. This proposal can be tested directly for those transitions for which domain boundaries carry topological charge. This is the case for annular Josephson tunnel Junctions (JTJs), for which the topological charge is the magnetic flux carried by a vortex 'fluxon' in the plane of the oxide layer between the two superconductors that make up the JTJ. In this article we shall explain how to devise experiments that test the KZ scenario and discuss the results to date.

However, before doing so, we should explain why we consider Josephson tunnel Junctions to be good candidates for testing the role of causality.

The basic ideas of Zurek and Kibble are straightforward and consist of two assumptions. We begin with the observation that, for the continuous transitions that we are considering, the adiabatic correlation length $\xi_{\rm ad}(T)$ of the order-parameter field diverges as $T$ approaches the critical temperature $T_c$. In reality, for a transition that is implemented in a finite time, the correlation length cannot become infinite as the temperature $T(t)$ passes through $T_c$ because causality requires that the field can only order itself at a finite speed. The first assumption made by Zurek and Kibble is that the maximum value that the physical correlation length $\xi$ can take is $\bar{\xi} = \xi_{\rm ad}(T(\bar{t}))$ for some appropriate time $\bar{t}$. Further, the time $\bar{t}$ is to be determined by simple causal arguments e.g. it corresponds to the time in which the causal horizon is large enough to enclose a single correlation length. Whether this is observable directly, or not, depends on the nature of the frustration of the order parameter as it moves from one groundstate to the next. In many cases the order parameters cannot 'untwist' at these domain boundaries because of conserved topological charges. The second KZ assumption is that, if the symmetry breaking in the transition permits topological defects e.g. vortices, then the separation $\bar{\xi}$ of such defects at the time of their production is $\bar{\xi} \approx \xi_{\rm ad}(T(\bar{t}))$.

Suppose that $T(0) = T_c$. The quench time (inverse quench rate) $\tau_Q$ is defined by

$$\frac{T_C}{\tau_Q} = -\frac{dT}{dt}\bigg|_{T=T_C}. \qquad (1)$$

Since $\xi_{\rm ad}(T(\bar{t}))$ depends on $\tau_Q$, these assumptions lead to a scaling behaviour for $\bar{\xi}$ of the form

$$\bar{\xi} \approx \xi_0 \left(\frac{\tau_Q}{\tau_0}\right)^\sigma. \qquad (2)$$

The scaling exponent $\sigma$ in (2) depends *only* on the critical indices of the system, and can be calculated in mean field (and perhaps better), whereas the parameters $\xi_0$ and $\tau_0$ have to be determined from the characteristics of the particular sample being tested. See [1–3] for greater details.

Several experiments have been performed to check the predictions of (2) in condensed matter systems, both in scaling behaviour and in normalisation, with varying degrees of success. The most obvious transitions that permit testing involve superfluids or superconductors, which we shall refer to in more detail later, since each has vortices as a natural topological defect.

The virtue of the prediction (2) is that it makes hardly any direct reference to the microscopic properties of the system. However, there are several potential problems. The first is that (2) does not, a priori, distinguish between defect-antidefect and defect-defect separation. It leads to a density of defects *plus* antidefects, and not to a density of topological charge. As a result, if there is defect-antidefect annihilation after the transition, we need additional information on the rate of this if we are to extrapolate back to the gross density at the time of their production.

Experimentally, this latter point has turned out to be a particular problem for testing (2) in quenches of $^4He$ from the fluid to superfluid phase. The standard experiments [4, 5] use a pressure quench to take the helium into the superfluid phase, and then use attenuation of second sound to infer the initial density of vortices produced in the irrotational system. This presupposes detailed knowledge of the rate of decay of vortex tangles with great accuracy, since the signal depends exponentially on the vortex density. The absence of any signal in the most recent experiment [5] is not a demonstration of the invalidity of (2), but more a reflection of an underestimation of the rate of vortex decay [6] and the effect of thermal fluctuations (or, equivalently, the importance of the Ginzburg regime) in a quench at approximately constant temperature [7]. The vortices seen in [4] are now understood as being created spuriously in the quench.

The situation is better with $^3He$, which can be elevated from superfluid to normal fluid by bombarding with slow neutrons, which can cause its break up into tritium with the release of enough energy to create hotspots into which vortices form on being cooled by the surrounding superfluid. The vortices formed can be observed directly with NMR [8], or indirectly by looking for an energy deficit [9]. Both experiments show good agreement with (2) in overall magnitude. However, because of the nuclear nature of the heating, the scaling exponent is not confirmed directly.

This is rectified to some extent by planar high-$T_c$ superconductors, except that here the experimental observable is conserved total flux through the superconductor i.e. the net topological charge or the number of Abrikosov vortices *minus* the number of antivortices. To distinguish between defects and antidefects properly requires more than just causality. Simple modeling suggests Gaussian behaviour at early times, which implies a random walk in phase along the superconductor boundary. As a result we would expect the variance in the flux to obey [2]

$$\Delta \Phi \approx \Phi_0 \left(\frac{C}{\bar{\xi}}\right)^{1/2} = \Phi_0 \left(\frac{C}{\xi_0}\right)^{1/2} \left(\frac{\tau_Q}{\tau_0}\right)^{-\sigma/2}. \tag{3}$$

High-$T_c$ superconductors are preferred because the smaller value of $\xi_0$ leads to a greater net flux. Although the first experiment [10] was null, because of low quench rates, a later experiment [11] saw flux spontaneously produced with scaling commensurate with (3), but at a low efficiency. However, with $\sigma$ expected to take the value of $1/4$, this one-eighth power behaviour of $\Delta \Phi$ with $\tau_Q$ is too difficult to establish experimentally with any accuracy.

There is another issue in that, for superconductors, there is a further mechanism [12] for producing spontaneous flux, in that the long wavelength modes of the magnetic field will fall out of equilibrium at the quench, giving rise to additional flux. Fortunately, for the experiment in hand this additional contribution is expected to be small.

The situation is simpler for annular JTJs, for which the defects are vortex flux tubes in the insulating oxide, in the plane of the two annuli of superconductors, the 'fluxons' of the model. Yet again, we are measuring the number of fluxons minus the number of antifluxons but, as we shall see, for a junction of circumference $C$, it is possible to interpolate between scaling behaviour of (3) for fast quenches and

$$\Delta \Phi \approx \Phi_0 \left(\frac{C}{\bar{\xi}}\right) = \Phi_0 \left(\frac{C}{\xi_0}\right) \left(\frac{\tau_Q}{\tau_0}\right)^{-\sigma} \tag{4}$$

for slow quenches with $\bar{\xi} < C$ where, for symmetric junctions, $\sigma$ is [13, 14], again, $1/4$. This gives us a chance of measuring $\sigma$. [The angle subtended by the oxide is so small that we need not consider the second mechanism mentioned above.] A first experiment by us [15, 16] confirmed the exponent $\sigma = 1/4$ of (4) within errors and a second (current) experiment looks to do the same. These experiments are the only ones for condensed matter systems that have been able, to date, to determine these exponents, whose values are critical to the notion that causality is the determinant in early domain formation.

Attempts to extend the KZ analysis to Type-I superconductors are intriguingly inclusive [17]. Experiments have been performed on liquid crystals [18] and rings of Josephson junctions [19] that are compatible with the ZK analysis without being in a position to demonstrate scaling behaviour. Further applications of the ZK analysis may arise in the experiments on vortex production in BEC that are currently being performed, and in two-gap superconductors and metallic hydrogen.

Of course, there have been several numerical simulations performed for time-dependent Landau-Ginzburg (TDLG) equations (e.g. see [20]) which, arguably, mimic dissipative systems like those above, to show the required scaling behaviour.

However, because of the generality of the ideas, the same scaling behaviour has been looked for (without success) in convection currents and (with success) in non-linear optics [21, 22]. We shall not consider other systems or simulations any further. For the remainder of this article we restrict ourselves to a description of JTJs from the viewpoint of causality and causal horizons and show how it is possible to derive and test the behaviours of (3) and (4) above.

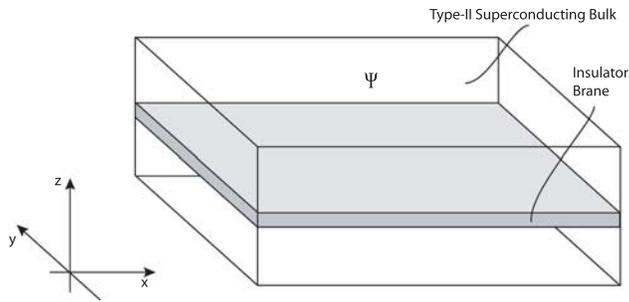

FIG. 1: A symmetric Josephson tunnel junction (JTJ). $\Psi$ is the common complex order parameter field.

## II. SINE-GORDON PLASMA MODES AND FLUXONS

Consider a Josephson junction in the form of a bulk superconductor separated by an insulating oxide (mem)brane as shown in Fig.1. [The nomenclature follows from the correspondence with cosmological field theory, which displays similar scaling behaviour, in which branes are worlds embedded in higher dimensional spaces on which our fields of interest live.] The effective order parameter field of the superconductor is $\Psi = \rho^{1/2} e^{i\phi}$, where $\rho$ is the density of Cooper pairs. The Landau-Ginzburg free energy due to $\Psi$ can be written as

$$F = F_{\text{bulk}} + F_{\text{brane}}, \tag{5}$$

where

$$F_{\text{bulk}} = \int d^3x \left( \frac{\hbar^2}{4m} |\nabla \Psi|^2 - \alpha |\Psi|^2 + \frac{\beta}{2} |\Psi|^4 \right) \tag{6}$$

and $F_{\text{brane}}$ is restricted to the $x - y$ plane as

$$F_{\text{brane}} = \int d^3x \, \delta(z) f_{\text{brane}} \tag{7}$$

with [25]

$$f_{\text{brane}} = a(|\Psi_+|^2 + |\Psi_-|^2) - b(\Psi_+^* \Psi_- + \Psi_-^* \Psi_+), \tag{8}$$

in which

$$\Psi_\pm = \lim_{z \to 0_\pm} \Psi(x,y,z) = \rho^{1/2} e^{i\phi_\pm}.$$

The groundstates of the bulk superconductor comprise the circle $|\Psi_0| = \rho_0^{1/2} = (\alpha/\beta)^{1/2}$ of Fig. 2. We assume continuity in $\rho$ across the oxide brane, but the second term in (8) allows for a discontinuity $\phi = \phi_1 - \phi_2$ in phase. The extremisation of $F$ leads to the equations [25]

$$0 = -\frac{\hbar^2}{4m} \nabla^2 \Psi - \alpha \Psi + \beta \Psi |\Psi|^2 \tag{9}$$

for the bulk, and

$$0 = \frac{\hbar^2}{4m} \frac{\partial \Psi_+}{\partial z} + a\Psi_+ - b\Psi_- \tag{10}$$

$$0 = -\frac{\hbar^2}{4m} \frac{\partial \Psi_-}{\partial z} + a\Psi_- - b\Psi_+ \tag{11}$$

for the brane.

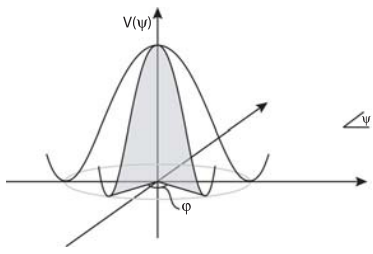

FIG. 2: The potential for the order-parameter field $\Psi$. $\phi$ is the gauge-invariant difference of $\Psi$ phases across the oxide brane.

The Cooper pair tunnelling current through the brane is

$$\vec{J} = -\frac{e}{m}\frac{i\hbar}{2}[\Psi^*\nabla\Psi - (\nabla\Psi^*)\Psi] \tag{12}$$

which, from (10) and (11), has transverse component

$$J_z = J_c \sin\phi, \tag{13}$$

the DC Josephson current, with $J_c \propto b$.

If we apply a DC potential $V$ across the brane, so as to shift the chemical potential by $\Delta\mu = 2eV$, the corresponding shift in phase $\phi \to \phi + \Delta\mu t/\hbar$ implies

$$\frac{\partial\phi}{\partial t} = \frac{2eV}{\hbar} \tag{14}$$

and leads to the AC Josephson current

$$J = J_c \sin(\phi_0 + 2eVt/\hbar).$$

The discussion above has not taken the electromagnetic field fluctuations into account, which require that we replace $\nabla$ by the covariant derivative $\nabla - 2ie\vec{A}$ in (6) (and add $B^2/8\pi$ to the free energy density). The end result is the familiar story; the Anderson-Higgs-Kibble mechanism shows how, in the interior of each of the bulk superconductors, the phases $\phi_\pm$ can be gauged away, to give a massive photon with finite (London) penetration depth. In addition there is a density field $\rho$, whose excitations describe a massive real scalar field (Higgs mode). However, in the brane the *difference* $\phi$ of the phases (the difference of the Goldstone modes) is gauge invariant, and cannot be eliminated.

The equations of motion for this Anderson plasma mode $\phi$, which only lives on the brane, are straightforward to derive. In the presence of the electromagnetic field the current equation (12) can be rewritten as

$$\nabla\phi = \frac{2e}{\hbar c}\left(\frac{mc}{2e^2\rho^2}\vec{J} + \vec{A}\right). \tag{15}$$

Viewed from the side, the junction of Fig. 1 has a magnetic depth $d_e \approx 2\lambda_L + d_{\text{ox}}$ (for $\lambda_L$ small enough) into which the magnetic field can penetrate, where $\lambda_L$ is the London length for each bulk superconductor, and $d_{\text{ox}}$ is the oxide thickness. Assuming no Abrikosov vortices in the bulk, applying (15) to the contour of Fig.3, and its counterpart along the $y$-axis leads to [23]

$$\frac{\partial\phi}{\partial x} = \frac{2e}{\hbar c}d_e H_y \tag{16}$$

$$\frac{\partial\phi}{\partial y} = -\frac{2e}{\hbar c}d_e H_x. \tag{17}$$

If we now insert the results of (16) and (17) into the transverse Maxwell equation

$$\frac{\partial H_y}{\partial x} - \frac{\partial H_x}{\partial y} = \frac{4\pi}{c}J_z + \frac{1}{c}\frac{\partial D_z}{\partial t}, \tag{18}$$

we find

$$\frac{\hbar c}{2ed_e}\left[\frac{\partial^2\phi}{\partial x^2} + \frac{\partial^2\phi}{\partial y^2}\right] = \frac{4\pi}{c}\left[J_c\sin\phi + \mathcal{C}\frac{\partial V}{\partial t}\right], \tag{19}$$

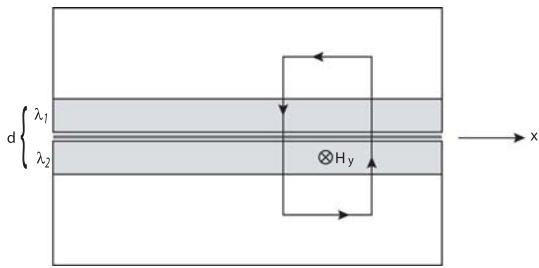

FIG. 3: The JTJ of Fig. 1 viewed from one side. The contour adopted to derive Eq.16 extends beyond the magnetic depth (shaded).

where $\mathcal{C}$ is the junction capacitance per unit area. On eliminating $V$ through (14), we finally arrive at the Sine-Gordon equation for the plasma mode $\phi$,

$$\frac{1}{\bar{c}^2}\frac{\partial^2 \phi}{\partial t^2} - \left(\frac{\partial^2 \phi}{\partial x^2} + \frac{\partial^2 \phi}{\partial y^2}\right) + \frac{1}{\lambda_J^2}\sin\phi = 0. \tag{20}$$

This is the equation of motion that would follow from extremising the three-dimensional space-time action

$$S_{SG} = \int d^3x \left[\frac{1}{2}\partial_\mu\phi\partial^\mu\phi + \frac{1}{\lambda_J^2}(1-\cos\phi)\right], \tag{21}$$

with $\partial_0 = (1/\bar{c})\partial/\partial t$. In (20) the speed of propagation of the plasma excitations is the Swihart velocity

$$\bar{c} = c\left(\frac{1}{4\pi \mathcal{C} d_e}\right)^{1/2}, \tag{22}$$

effectively the speed of light in the brane. The Josephson coherence length of the plasma modes is

$$\lambda_J = \left(\frac{\hbar c^2}{8\pi e d_e J_c}\right)^{1/2}, \tag{23}$$

where, more generally, for finite thickness superconducting electrodes whose thickness id $d_s$

$$d_e = d_{ox} + 2\lambda_L \tanh\frac{d_s}{2\lambda_L}.$$

Consider now the quenching of the system from above its critical temperature to some temperature significantly below $T_c$. At the end of the transition the field typically finds itself in one of the many degenerate vacua of the potential of (21), as shown in Fig.4. The plasma oscillations take place around a local minimum. However, if we follow the phase around the boundary of the junction (e.g. around the boundary of the insulating oxide in Fig.1, treating the junction as a finite system) we shall find 'kinks' in $\phi$ where it changes value by $\pm 2\pi$ over a length $O(\lambda_J)$. They correspond to jumping from one minimum to an adjacent one in Fig.4. When this happens the Josephson current changes direction as we pass through $\Delta\phi = \pi$, showing the presence of a unit of magnetic flux leaving the oxide layer. This is a Josephson fluxon, a solution to the Sine-Gordon equation (20) which, when static, has the form

$$\phi_\pm(x,0) = 4\arctan\exp(\pm x/\lambda_J). \tag{24}$$

In fact, (20) is oversimplified, since $J_z$ contains secondary terms in addition to $J_c\sin\phi$, in particular a quasi-particle tunneling current proportional to $V$, and thereby to $\partial\phi/\partial t$, which introduces dissipation into the Sine-Gordon equation. However, such terms do not stop the Swihart velocity being the maximum speed at which a profile of this form or, indeed, of any form can move.

Further, as we continue around the boundary, we expect a random walk in phase, as is Fig.4.

### III. CRITICAL BEHAVIOUR

In the vicinity of the second-order conductor-superconductor transition at temperature $T_c$, we adopt the Landau-Ginzburg form for (6), in which $\alpha$ is replaced by

$$\alpha(T) = \alpha_0\left(1 - \frac{T}{T_c}\right). \tag{25}$$

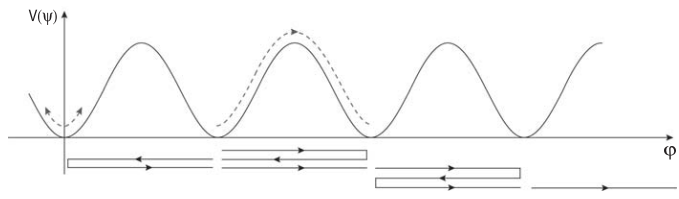

FIG. 4: The SG potential $V(\phi) = 1 - \cos\phi$. Plasma oscillations are confined to a single minimum, whereas the fluxon corresponds to the transition from one minimum to the next. The random walk in phase $\phi$ expected along annuli of large circumference is displayed under the potential.

This goes together with a temperature-dependent Josephson coherence length and a temperature-dependent Swihart velocity. The connection is not direct, and relies on the relationship

$$J_c(T) = \frac{\pi}{2} \frac{\Delta(T)}{e\rho_N} \tanh \frac{\Delta(T)}{2k_B T} \qquad (26)$$

between $J_c(T)$ and the superconducting gap energy $\Delta(T)$, that can only be obtained from a fundamental BCS theory of electrons, which has the free energy of (6) and (8) as an effective representation. In (26) $\rho_N$ is the JTJ normal resistance per unit area. The temperature dependence of the gap energy $\Delta$ in a strong-coupling superconductor satisfies

$$\frac{\Delta(T)}{\Delta(0)} = \tanh \frac{\Delta(T)}{\Delta(0)} \frac{T_c}{T}. \qquad (27)$$

and varies steeply near $T_c$ as

$$\Delta(T) \simeq 1.8\,\Delta(0)\left(1 - \frac{T}{T_c}\right)^{1/2}. \qquad (28)$$

Introducing the dimensionless quantity $\alpha = 1.8\Delta(0)/k_B T_C$ ($\Delta(0)$ and $J_c(0)$ denote the respective values at $T = 0$) whose typical value is between 3 and 5, enables us to write $J_c(T)$ as

$$J_c(T) \simeq \alpha J_c(0)\left(1 - \frac{T}{T_c}\right). \qquad (29)$$

Putting (26) and (28) together gives the behaviour, in the vicinity of the transition,

$$\lambda_J(T) = \xi_0\left(1 - \frac{T}{T_c}\right)^{-1/2}, \qquad (30)$$

where

$$\xi_0 = \sqrt{\frac{\hbar}{2e\mu_0 d_s \alpha J_c(0)}}. \qquad (31)$$

As expected $\lambda_J(T)$ diverges as $T \to T_c$ from below.

On the other hand, for a finite electrode thickness tunnel junction, the Swihart velocity of (22) takes the form[23]

$$\bar{c}(T) = c_0\sqrt{d_{ox}/\epsilon_r d_i(T)}, \qquad (32)$$

where

$$d_i(T) = d_{ox} + 2\lambda_L(t) \coth \frac{d_s}{\lambda_L(T)} \simeq \frac{2\lambda_L^2(0)}{d_s}\left(1 - \frac{T}{T_c}\right)^{-1},$$

near the transition. Thus $\bar{c}(T)$ shows critical slowing down at the transition, as

$$\bar{c}(T) = \bar{c}_0\left(1 - \frac{T}{T_c}\right)^{1/2}, \qquad (33)$$

where $\bar{c}_0 = c_0\sqrt{d_s d_{ox}/2\epsilon_r \lambda_L^2(0)}$. From $\xi_0$ and $\bar{c}_0$ we can construct

$$\tau_0 = \xi_0/\bar{c}_0,$$

which is the timescale $\tau_0$ of (2).

We complete this preliminary discussion of JTJs by considering *non-symmetric* junctions in which the superconductors, 1 and 2, above and below the brane in Fig.1 now have different critical temperatures $T_{c2} > T_{c1}$. The order parameter remains the plasma mode $\phi = \phi_1 - \phi_2$, the difference in the Goldstone modes of the two superconductors. Fluxons only appear at temperatures $T < T_{c1}$, from which we measure our time $t$. At this time

$$\Delta_2(T_{c1}) \simeq 1.8\,\Delta_2(0)\left(1 - \frac{T_{c1}}{T_{c2}}\right)^{1/2},$$

and, otherwise,

$$\Delta_1(T) \simeq 1.8\,\Delta_1(0)\left(1 - \frac{T}{T_{c1}}\right)^{1/2}. \tag{34}$$

The critical Josephson current density $J'_c(T)$ for a non-symmetric JTJ, being proportional to $\Delta_1(T)\Delta_2(T)$ [23], behaves just after the transition as

$$J'_c(T) \approx \left(1 - \frac{T_{c1}}{T_{c2}}\right)^{1/2} \alpha' J'_c(0)\left(1 - \frac{T}{T_{c1}}\right)^{1/2}, \tag{35}$$

where $J'_c(0) = \pi\Delta_1(0)\Delta_2(0)/[\Delta_1(0) + \Delta_2(0)]e\rho_N$, and $\alpha' = [\Delta_1(0) + \Delta_2(0)]/k_B T_{c,1}$, provided $\Delta_2(T_{c,1}) \ll 2\pi k_B T_{c,1}$. The crucial difference between (35) and (29) is in the critical index. Near $T = T_{c1}$, we now find

$$\lambda_J(T) = \xi_0\left(1 - \frac{T_{c1}}{T_{c2}}\right)^{-1/4}\left(1 - \frac{T}{T_{c1}}\right)^{-1/4}, \tag{36}$$

where $\xi_0$ is as before, since $J'_c(0)$ is indistinguishable from $J_c(0)$ and $\alpha'$ is comparable to $\alpha$.

Finally, let us consider the case, as in Fig. 1, in which the insulating oxide brane is replaced by a *total* insulator, with critical Josephson current $J_c = 0$. The phase around the upper edge of the insulator and the phase around the lower edge are now totally independent. However, in the same mean field approximation that we have used above, it follows from (25) that the correlation length in phase has the *same* form as (30) for the symmetric JTJ,

$$\lambda(T) = \xi_s\left(1 - \frac{T}{T_c}\right)^{-1/2}, \tag{37}$$

where $\xi_s$ depends on the nature of the superconductor (see [1, 2]). Similarly, there is a characteristic relaxation time-scale $\tau_s$ such that the speed at which the field can order itself is

$$c_s(T) = c_s\left(1 - \frac{T}{T_c}\right)^{1/2}, \tag{38}$$

where $c_s = \xi_s/\tau_s$. [Note that lower-case $s$ denotes *superconductor* and not upper-case *Swihart*]. Again there is the same slowing down as in (33). The critical temperature of the superconductor is approximately that of the JTJ and we have not discriminated between them.

## IV. CAUSAL BOUNDS AND SCALING BEHAVIOUR

Although the correlation length $\xi_{\text{ad}}(T)$ of the $\phi$ field diverges as we approach the critical temperature from below we know that causality prevents the JTJ from having an infinite correlation length when we cool the junction through its critical temperature.

For the JTJ of Fig.1, the behaviour of the flux lines within the $x - y$ plane of the oxide is unclear. [We stress that, as a first approximation, we assume that there are no Abrikosov vortices in the $z$- direction]. Henceforth we restrict ourselves to the simpler case of a long thin JTJ, whose extension in the $y$-direction of the junction is $w \ll \lambda_J$, for which the fluxons have topological charge $\Delta\phi/2\pi = \pm 1$. In fact, for the purpose of measuring flux it is convenient to bend this into an annulus in which the insulating oxide brane lies in the plane of the annulus and our subsequent discussion is restricted to such annuli. To give a sense of scale we shall consider annuli of circumference $500 - 1000 \mu$m in perimeter and a few $\mu$m in width. If we pass through a transition into the superconducting phase rapidly enough we shall expect to see fluxons produced spontaneously along its perimeter as the field organises itself after the quench.

We now have to decide what speeds define causality for the $\phi$ field and, further, how to define the causal horizon. We have seen that there are two possibilities; the Swihart velocity $\bar{c}$ and the velocity $c_s$ for phase change in an individual superconductor. We shall see that, although (30) and (37) have the same critical indices (for symmetric superconductors), there are very different signatures.

Let us first assume that causality is determined by the Swihart velocity. In the vicinity of $T_c$ ($t > 0$) we have, from (1),

$$\left(1 - \frac{T}{T_c}\right) \approx -\frac{t}{\tau_Q} \tag{39}$$

whereby, for small $t$,

$$\lambda_J(t) = \lambda_J(T(t)) = \xi_0 \left(\frac{\tau_Q}{t}\right)^{1/2}, \tag{40}$$

whereas

$$\bar{c}(t) = \bar{c}_0 \left(\frac{t}{\tau_Q}\right)^{1/2}. \tag{41}$$

As we hinted in the introductory paragraphs of this article, there is no unique way to determine $\bar{t}$ [1–3]. In fact, there are two markedly different ways to proceed, each with its own variants. They all agree approximately for simple cases, but JTJs are not simple and it is worthwhile to briefly rehearse them. The original KZ proposals were based on the assumption that $\bar{t}$ is determined by the behaviour of the system for *negative* $t$ (i.e prior to the transition having begun). Two possible ways to defend this are to argue that the system does its best to adjust to the quench, but freezes in at the time $t = -\bar{t}$ and unfreezes at time $t = \bar{t}$, after which the adiabatic approximation is adequate. From this viewpoint $t = -\bar{t}$ can be thought of as the time a) when the relaxation time for the long wavelength modes at time $\bar{t}$ is, itself, $\bar{t}$ or b) when the correlation length $\xi_{\text{ad}}(T(t))$ is changing as fast as possible. i.e.

$$\frac{d\xi_{\text{ad}}(-\bar{t})}{dt} = \bar{c}(-\bar{t}). \tag{42}$$

These ideas are not applicable to JTJs, since the Josephson effect does not exist prior to the transition, and we need approaches that concentrate on the appearance of defects after the transition has begun.

Fortunately, the latter approach permits generalisation to

$$\frac{d\xi_{\text{ad}}(\bar{t})}{dt} = -\bar{c}(\bar{t}), \tag{43}$$

as the first time *after* the quench that it makes sense to identify defects, since (43) does not really rely on an 'impulse' phase in which the system is frozen. Naturally, both (42) and (43) give the same result.

A more intuitive approach is to estimate the causal horizon directly. Again suppose that $\bar{c}(t)$ determines the causal horizon for the ordering of the plasma mode from time $t = 0$ onwards. Then, at time $t$ the length of junction in which the order parameter $\phi$ can be correlated is, as a first guess,

$$\xi_c(t) \approx 2 \int_0^t ds\, \bar{c}(s). \tag{44}$$

Our horizon bound for the earliest time $\bar{t}$ that we can see fluxons is when $\xi_c(t)$ becomes larger than the Josephson coherence length $\lambda_J(t)$. That is, the causal horizon is big enough to hold a classical fluxon,

$$\xi_c(\bar{t}) = \lambda_J(\bar{t}). \tag{45}$$

Equivalently, this is the first time that the causal horizon is as large as the instantaneous correlation length. Up to factors close enough to unity to be taken as unity at our level of approximation, (45) gives the same result as (43). For the symmetric JTJs above simple algebra shows that

$$\bar{t} \approx \sqrt{\tau_0 \tau_Q}, \tag{46}$$

where $\tau_0 = \xi_0/\bar{c}_0$.

If we identify $\xi_{\rm ad}(t)$ with $\lambda_J(t)$, then

$$\bar{\xi}_J = \lambda_J(\bar{t}) \approx \xi_0 \left(\frac{\tau_Q}{\tau_0}\right)^{1/4} \tag{47}$$

is the $\phi$ correlation length at the time that fluxons are formed. We stress that (47) is only an order of magnitude estimate, or better.

A priori it may seem unreasonable that the density of flux should be determined by the speed of light along the insulator layer. One might argue that it is the way the flux is formed in the magnetic depth of the superconducting bulk that is important. If that is so, we have a situation in which, approximately, it is the behaviour of the individual superconductors that matter. Insofar that $\phi_+$ and $\phi_-$ are independent at the time that fluxons can be counted, we can repeat the calculations of Zurek for a single superconducting annulus [2] (perhaps taking $\Delta\phi = \sqrt{2}\Delta\phi_\pm$).

From the identical critical behaviour of superconductors and symmetric JTJs we find, again, that

$$\bar{t} \approx \sqrt{\tau_s \tau_Q}. \tag{48}$$

If we identify $\xi_{\rm ad}(t)$ with $\lambda(t)$, then

$$\bar{\xi}_s = \lambda(\bar{t}) \approx \xi_s \left(\frac{\tau_Q}{\tau_s}\right)^{1/4} \tag{49}$$

is the $\phi$ correlation length at the time that fluxons are formed, rather than (47).

Despite their differences in normalisation, (47) and (49) give us the important result that, for symmetric JTJs, the fluxon separation $\bar{\xi}$ at the time of their formation satisfies the scaling behaviour

$$\bar{\xi} = O(\tau_Q^{1/4}), \tag{50}$$

whatever the mechanism. This is the key result of our analysis and it is the confirmation of this exponent that our experiments address first.

On the other hand, to determine which is the relevant velocity requires understanding the normalisation. Given the approximate nature of the bounds (47) and (49) this can be confusing. Empirically, $\tau_s \lesssim \tau_0$ for the JTJs that we have considered (in each case a fraction of a picosecond). However, whereas $\tau_0$ and $\tau_s$ are comparable, $\xi_s \ll \xi_0$, in general. Typically, $\xi_s$ is tenths of microns, whereas $\xi_0$ is of order 10 microns. Thus we expect up to an order of magnitude greater spontaneous flux after a random walk in phase along an annulus if the flux formation is determined by the individual superconductors rather than by the Swihart velocity. The effect is even more dramatic when (4) is appropriate, with the possibility of two orders of magnitude difference in the likelihood of observing a single fluxon in small annuli, depending on the mechanism.

There are two obvious ways to clarify the situation, that rely on comparison, rather than absolute scale. The first is to use samples in which the superconductors are identical, but in which the critical current density $J_c$ is strongly different. This can be easily achieved experimentally by controlling the tunnel barrier thickness, i.e. the Josephson coupling. Since $\xi_0$ is larger for smaller $J_c$, smaller $J_c$ gives less spontaneous flux if the Swihart velocity is the determining factor. If, however, it is the bulk superconductors that set the scale, then reducing the critical current will not reduce the spontaneous flux, at a first approximation.

The second way is to use non-symmetric superconductors, for which (36) and (38) follow. We have shown that, if the Swihart velocity determines causality, then

$$\bar{t} = \tau_0^{4/7} \tau_Q^{3/7} \left(1 - \frac{T_{c1}}{T_{c2}}\right)^{-1/7} s.$$

In turn,

$$\bar{\xi}'_J = \lambda_J(\bar{t}) \simeq \xi_0 \left(1 - \frac{T_{c1}}{T_{c2}}\right)^{-1/4} \left(\frac{\tau_Q}{\tau_0}\right)^{1/7} \tag{51}$$

In practice, this gives a $\bar{\xi}'_J$ an order of magnitude smaller than $\bar{\xi}_J$ of (47) and an order larger than $\bar{\xi}_s$ of (49), giving an intermediate amount of spontaneous flux. However, the scaling exponent in (51) is so small as to make this flux almost independent of quench rate. This is in contrast to the behaviour that we would expect if it was the bulk superconductors that controlled the flux, since we would expect the result of (49) to survive, qualitatively. We have yet to perform experiments on non-symmetric JTJs and will not comment further on them.

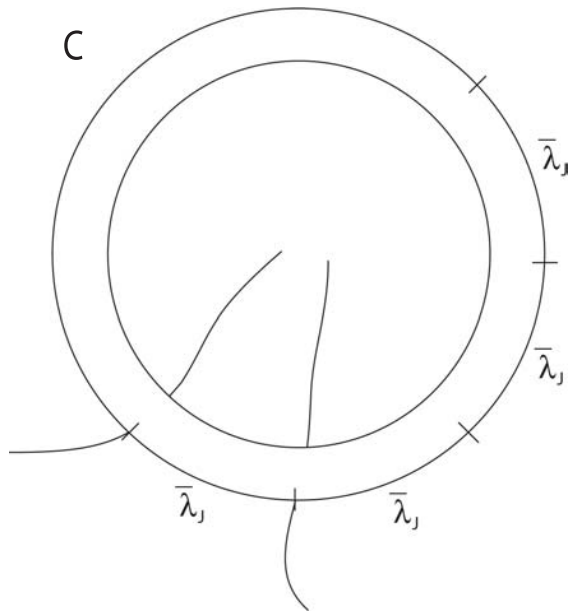

FIG. 5: Fluxons in a large annulus. To be specific we have assumed that it is the Swihart velocity that has set the correlation length.

If we assume that the results above are unaffected by the periodic boundary conditions on our annulus of circumference $C \gg \bar{\xi}$ then we would expect the situation as given in Fig.5 on quenching the annulus; on average there will be one fluxon or antifluxon per every correlation length $\bar{\xi}$ (in the case displayed $\bar{\xi} = \bar{\lambda}_J$).

Experimentally we can only measure the net flux through the annulus. On average it will be zero, but assuming a random walk in phase along the annulus suggests that the rms deviation in the net flux will be

$$\Delta \Phi \approx f \left( \frac{C}{2\pi\bar{\xi}} \right)^{1/2} \Phi_0, \tag{52}$$

as we had anticipated in (3). We have now included an efficiency factor $f$ for producing defects, that determines to what extent the causal bounds are saturated. We have explicitly removed a geometrical factor $(2\pi)^{-1/2}$ in $f$ because a fluxon requires $\Delta\phi = 2\pi$. We expect $f$ to be unity to an order of magnitude or better.

In practice, as we shall show in greater detail, $\Delta\Phi/\Phi_0 < 1$ in our experiments. However, once $C/\bar{\xi}$ is somewhat less than unity a random walk in phase is inappropriate. Instead, we estimate that the probability $P_1$ of observing a *single* fluxon (or antifluxon) is

$$P_1 \approx \frac{f_1}{2\pi} \left( \frac{C}{\bar{\xi}} \right) = \frac{f_1}{2\pi} \frac{C}{\xi_0} \left( \frac{\tau_Q}{\tau_0} \right)^{-1/4}, \tag{53}$$

where $f_1$ is an efficiency factor. [To reduce the variety of suffices we have used $\xi_0$ and $\tau_0$ to be the fundamental distance and time scales, whether determined by the Swihart velocity or the individual superconductors.] Again, we expect $f_1$ to be unity to an order of magnitude. Assuming that, if the probability of finding one fluxon is small, the probability of finding two fluxons and an antifluxon (or vice-versa) are negligibly small at the level of the somewhat better than order-of-magnitude predictions that we can make, then we can convert (53) into the result that the probability for observing a single unit of flux $\Phi_0$ is $n_1 = P_1$.

## V. EXPERIMENTS WITH ANNULAR JTJS

To date, we and our colleagues are the only group to have performed experiments on JTJs on the lines suggested earlier. In 2001/2 the first experiment was performed successfully. Details can be found in the literature [15, 16], for which we shall provide a summary later. This first experiment has lead to a further sequence of experiments by us that are currently being performed. Although we have preliminary data, it would be premature to comment on them in any detail [29].

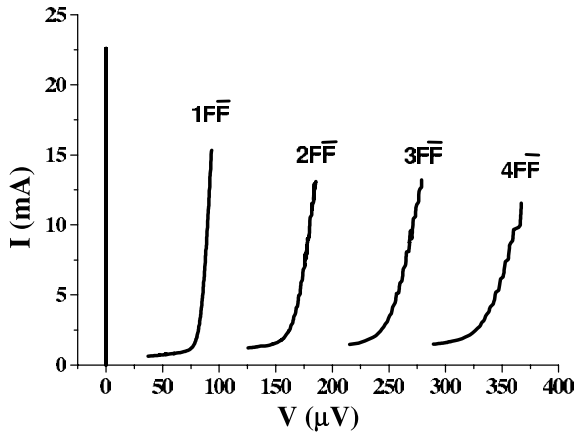

FIG. 6: The measured current-voltage characteristics of an AJTJ without trapped fluxons. For each current branch the number of fluxon-antifluxon pairs $F\bar{F}$ is indicated.

### A. Experimental Methods

To begin with some generalities, the JTJs that we have used are high quality $Nb/Al - Al_{ox}/Nb$ JTJs fabricated on $0.5mm$ thick silicon substrates using the trilayer technique in which the junction is realized in the window opened in a $SiO$ insulator layer. Details of the fabrication process can be found in Ref.[30]. The samples were fabricated at the Superconducting Electronics Laboratory of the Institute of Radioengineering & Electronics of the Russian Academy of Science in Moscow, while the measurements were carried out at the Physics Department of the Danish Technical University in Lyngby. All the experiments performed to date have been on JTJs with a mean circumference $C = 500\,\mu m$ and a width $\Delta r = 4\,\mu m$. JTJs with larger circumferences have been fabricated, but we have yet to use them. The high quality of the samples was inferred from the I-V characteristic at $T = 4.2\,K$ by checking that the subgap current $I_{sg}$ at $2\,mV$ was small compared to the current rise $\Delta I_g$ in the quasiparticle current at the gap voltage $V_g$. The symmetry of the junctions is assured by the absence of a logarithmic singularity in the IVCs at low voltages and the linear temperature dependence of the critical current as the temperature $T$ approached the critical temperature $T_C$.

Firstly, there is the question of how we identify fluxons since, when they are produced, they are static. To see them it is necessary to feed a bias current to the annular JTJ, whereupon the fluxons move as magnetic dipoles with a profile obtained by substituting $(x - ut)\sqrt{1 - (u/\bar{c})^2}$ for $x$ in (24) where $|u| < \bar{c}$, whose maximum speed is the Swihart velocity. As a result, they leave a clear signature on the junction current-voltage characteristic (IVC). Fluxons having different topological charge $n = \pm 1$ travel in opposite directions. Quantitatively, if a fluxon travels around an annular JTJ having circumference $C$ with a constant speed $v < c(T)$, then an average voltage $V = \Phi_0 v/C$ develops across the junction. By changing the bias current through the barrier the voltage drop changes and a new branch appears on the junction IVC. When $N$ fluxons travel around an annular JTJ, the junction voltage is $V = N\Phi_0 C/v$. In this expression $N$ is the total number of travelling fluxons and can be larger than the winding number $n$ if $F\bar{F}$ pairs are travelling around the annulus. Therefore, we can count the number of travelling fluxons by simply measuring the voltage across the annular JTJ.

To show this, Figs.6 and 7 represent the IVC of the same annular JTJ with no fluxon trapped and with one fluxon trapped, respectively.

We note that with no trapped fluxons the zero voltage current is very large. In the other case the supercurrent is rather small (ideally zero) and large current branches can be observed at finite voltages corresponding to the fluxon and, possibly, $F\bar{F}$ pairs travelling around the junction.

Secondly, there is the question of how we measure $\tau_Q$. This again is straightforward, in principle, and uses the observation due to Thouless [26] that the junction itself acts as a thermometer. The analytical expression for the superconductor gap energy $\Delta(T)$,

$$\frac{\Delta(T)}{\Delta(0)} = \tanh\frac{\Delta(T)}{\Delta(0)}\frac{T_c}{T}, \tag{54}$$

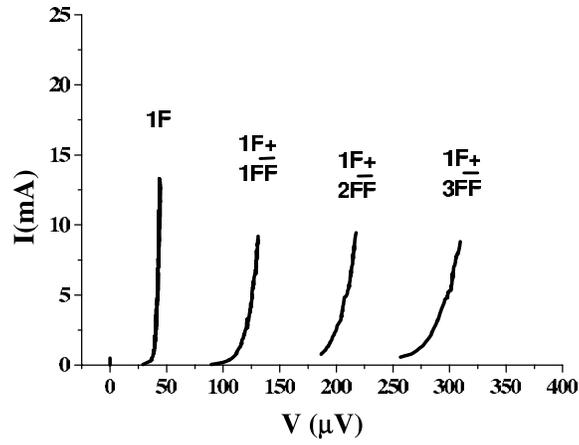

FIG. 7: The current-voltage characteristics of the same annular JTJ as in Fig.6 with one trapped fluxon and fluxon-antifluxon pairs.

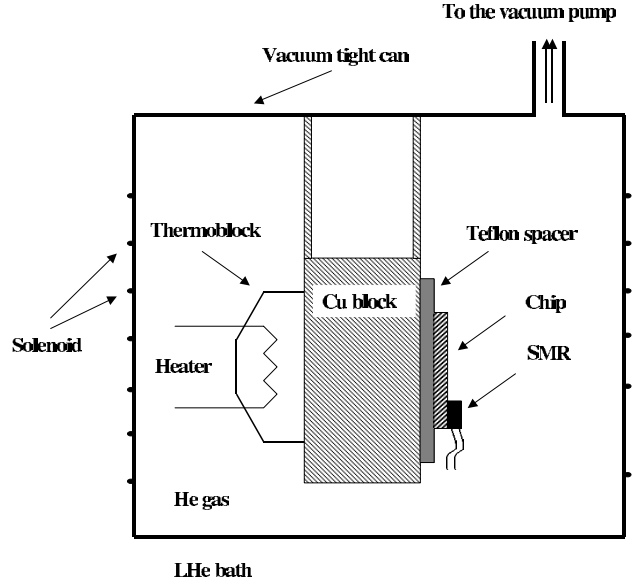

FIG. 8: Sketch (not to scale) of the cryogenic insert developed to perform the junction thermal cycles over slow and fast timescales

also applies to the junction gap voltage $V_g$ that is proportional to $\Delta$. In this way we derive temperature profiles for the thermal cycles, fitting our data with a thermal relaxation equation that describes the cooling system.

To be specific, let us consider our first experiment briefly. More details can be found in [15, 16] and conference proceedings [28]. The experimental setup is schematically shown in Fig.8. A massive $Cu$ block held to the sample holder was used to increase the system thermal capacity. The chip was mounted on one side of this block. On the other side, a thermoblock was mounted consisting of a $50\,\Omega$ carbon resistor and two thermometers in order to measure and to, if necessary, stabilize the $Cu$ block temperature. Finally a small sized $100\,\Omega$ surface mounted resistor (SMR), was kept in good thermal contact with the chip.

This system allowed us to perform the sample quenching over two quite different time scales. By means of the resistor in the thermoblock, a long time scale was achieved by heating the chip through the $Cu$ block; on the contrary, a short current pulse through the SMR on the chip, attained much short thermal cycles. The total system was kept in a vacuum-tight can immersed in the $LHe$ bath at $He$ gas pipeline pressure. By changing the exchange gas pressure and using the two techniques, the quenching time can be changed over a quite large range from tenths to tens of seconds. The disadvantage with this system is that we have no quenches of intermediate length between the two regimes.

We are only interested to the cooling process, and we use (54) to fit our data by a simple thermal relaxation

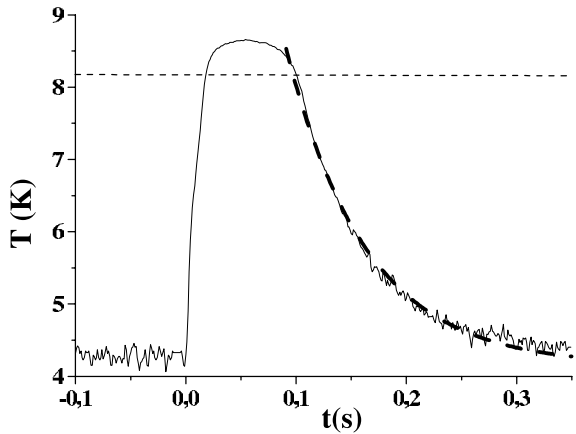

FIG. 9: Time dependence of the junction temperature (only reliable below the horizontal dashed line) during a slow thermal cycle, as determined from (54). The thick dashed line is the best fitting curve to (55).

equation:

$$T(t) = T_{fin} + (T_{in} - T_{fin}) \exp\left(-\frac{t-t_0}{\tau}\right) \tag{55}$$

with only two fitting parameters $t_0$ and $\tau$, and $T_{in}$ and $T_{fin}$ fixed at 8.95 and 4.15 $K$, respectively. In Eq.55 $t_0$ is the time at which $T = T_{in} = T_c$ and $\tau$ is the relaxation time which sets the cooling time scale. The fitting curve for a fast quench is shown by a thick dashed line in Fig.9, corresponding to a thermal relaxation time $\tau$ of $0.073\,s$. The dashed horizontal line indicates the temperature threshold below which the temperature time dependence can be reliably accounted for by our measured data. From the definition (1) for $\tau_Q$ it follows that $\tau_Q = \tau T_C/(T_{in} - T_{fin})$. This permits a measurement of $\tau_Q$ with 5% accuracy.

The experiment currently being performed [29] differs from the first in having a faster and more reliable heating system achieved with the integration of a resistive element. In stead of the $Cu$ block, heating is obtained by integrating a resistive element on the chip containing the JTJ.

Quenching experiments are carried out in a double $\mu$-metal shielded cryostat and the transitions from the normal to the superconducting states are performed with no current flowing in the heaters and the thermometers. Furthermore, the heat supplied to the sample is such that the maximum temperature reached by the junction is made slightly larger than its critical temperature, say at about $10\,K$, in order to make sure that also the bulk electrode critical temperature is exceeded. For each value of the quenching time, in order to estimate the trapping probability, we have carried out several hundred thermal cycles. At the end of each cycle the junction IVC is inspected in order to ascertain the possible spontaneous trapping of one or more fluxons.

As we shall see later, the annular JTJs are such that the probability of finding a single fluxon is less than unity. In the following we will focus our attention only on the probability $P_1$ to trap just one fluxon, although a few times we find clear evidence of two and, more seldom, three homopolar fluxons spontaneously trapped during the N-S transition. However, these events are too rare to be statistically significant. Experimentally, we define $P_1$ as the fraction of times in which at the end of the thermal cycle the junction IVC looks like Fig.7, i.e. with a tiny critical current and a large first ZFS. Our definition of $P_1$ is reasonable as far as the chance to trap two fluxons is negligible.

### B. Experimental Results.

As we have stressed, our main prediction is that, if causal horizons determine spontaneous fluxon formation then, providing $P_1 < 1$ (of which more later), $P_1$ has the allometric form

$$P_1 = a\,\tau_Q^{-b}, \tag{56}$$

where $b = 0.25$.

Many samples were measured in the first experiment. The results for Sample A are given in Fig. 10. The Swihart velocity $c_0$ has the value $c_0 = 1.4 \times 10^7 m/sec$. As expected, faster quenches produce more flux. Although the points are quite scattered, if we attempt to fit the data with the form (56) the best fit for the exponent is

$$b = 0.27 \pm 0.05,$$

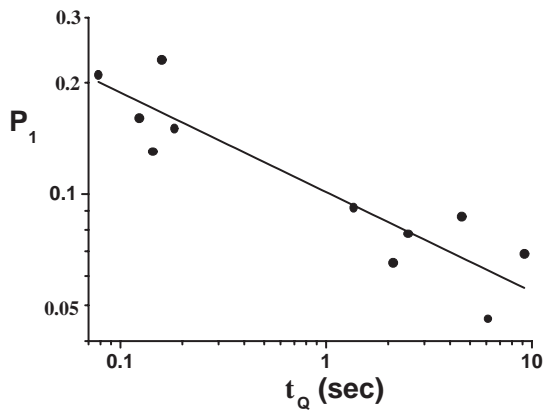

FIG. 10: Log-log plot of the probability $P_1$ to trap one fluxon versus $\tau_Q$. The solid line is the best fitting curve, assuming a power law dependence as in (53)

in remarkable agreement with our prediction.

The geometrical and electrical (at $4.2\,K$) parameters of Sample A and another sample, Sample B, are listed in Table I. They are chosen to agree in size and in the nature of their bulk superconductors, but differ strongly in their critical current densities.

TABLE I: The samples

| Sample | A | B |
|---|---|---|
| Mean circumference $C(\mu m)$ | 500 | 500 |
| Width $\Delta r(\mu m)$ | 4 | 4 |
| Zero field critical current $I_o(mA)$ | 33 | 2.5 |
| Maximum critical current $I_{\max}(mA)$ | 39 | 2.7 |
| Gap quasiparticle current step $\Delta I_g(mA)$ | 88 | 5.2 |
| $I_{\max}/\Delta I_g$ | 0.45 | 0.52 |
| Critical current density $J_c(A/cm^2)$ | 3050 | 180 |
| Josephson length $\lambda_J(\mu m)$ | 6.9 | 28 |
| Normalized mean circumference $C/\lambda_J$ | 72 | 18 |
| Quality factor $V_m(mV)$ | 49 | 63 |
| Normal resistance $R_N(m\Omega)$ | 36 | 610 |
| ZFS1 asymptotic voltage $(\mu V)$ | 51 | 53 |

Similar measurements to those for Sample A have been carried out for sample B. If spontaneous flux creation is determined by the bulk superconductors we would expect similar results for Sample B. If it is the Swihart velocity and $\bar\lambda_J$ which determines correlation lengths then we would expect to see much less fluxon creation in Sample B. This is due to a much smaller normalized length ($C$ measured in units of $\bar\lambda_J$) which, according to (53), translates in a expected probability $P_1$, for a given $\tau_Q$, about four times smaller. In fact, many less fluxons were seen. Unfortunately, although compatible with (53), the results were affected by a data scattering even larger than that found for sample A, since the probability of seeing a fluxon is far too small to get statistically significant data in reasonable times. [Each point on Fig. 10 corresponds to 300 thermal cycles.] However the roughly measured probability $P_1$ of one fluxon every 50-100 attempts is in fairly good agreement with the expected value.

For the coefficient $a$ we found the best fitting value of $0.1 \pm 10\%$ ($\tau_Q$ in seconds). This is to be compared with the predicted value of $f_1 C \tau_0^{1/4}/2\pi\xi_0$. Sample A had a circumference $C = 500\,\mu m$. Its effective superconductor thickness was $d_s \approx 250\,nm$. At the final temperature $T_{fin} = 4.2\,K$, the critical current density was $J_c(T_{fin}) = 3050\,A/cm^2$ and the Josephson length was $\lambda_J(T_{fin}) = 6.9\,\mu m$. From this, and $c_0$ given earlier, we infer that $\xi_0 \approx 3.8\,\mu m$ and $\tau_0 \approx 0.17\,ps$. This then gives $f_1 C \tau_0^{1/4}/2\pi\xi_0 \approx (f_1/2\pi)0.08\,s^{1/4}$, in agreement with the experimental value of $a$ if

$f_1 = 2\pi$. Given the fact that the genesis of the $2\pi$ factor in not robust, and we only expect agreement in overall normalization to somewhat better than an order of magnitude level this is further confirmation of the role played by the Swihart velocity.

In the second experiment, for which measurements are currently being made, we have little to say beyond the fact that the allometric form (56) provides a good representation for the slower quenches of Fig. 10 with $b \approx 0.25$ but that, provisionally there seems to be a deficit of fluxons at faster quenches. Details will be given elsewhere [29].

We conclude with a comment about the transition from the linear regime of (53) to the random phase regime of (52). Whether it is relevant to the new experiment or not, of itself it will lead to relative deficiency of flux. A first guess would suggest that the transition would occur when $P_1 \approx 1$, but simple diffusion models for the phase of a complex field suggest [27] that the transition could arise for values of $P_1$ as low as $P_1$ lying between $1/\pi$ and $1/2\pi$. These values were only just achieved in the first experiment, and we wait on the new experiments to clarify this further.

## VI. CONCLUSION

In this article we have examined the consequences of assuming that Josephson Tunnel Junctions change as fast as they can when quenched through the N-S transition. If this is the case for symmetric JTJs there is a clear signal; the likelihood of seeing a single fluxon in a quench with quench time $\tau_Q$ will scale as

$$P_1 = a\,\tau_Q^{-b}, \qquad (57)$$

where $b = 0.25$, provided $P_1 < 1$ is suitably small. Experiments performed to date confirm this value of $b$, with some qualifications. The coefficient $a$ depends on the details of how causality is implemented. If this is through the Swihart velocity, then $a$ is relatively small, increasing as the Josephson critical current increases. If it is through the independent superconductors then $a$ is relatively large, and independent of the critical current. The only complete experiment to date supports the former strongly, but new experiments are in hand.

This hypothesis of the importance of causality in determining the domain structure and thereby, the density of topological defects, after a continuous transition is due to Zurek and Kibble. Kibble proposed that this was also applicable to the early universe, with cosmological implications for the formation of topological defects. The possibility exists that we have our first sighting of a cosmic string [31], a field vortex that is the ubiquitous defect of all simple extensions of the Standard Model for the unification of fundamental forces.

This work was, in part, supported by the COSLAB programme of the European Science Foundation.